\documentclass[twocolumn,pra,aps]{revtex4-1}

\usepackage{mathptmx}
\usepackage{subfigure}
\usepackage{psfrag,graphicx}
\usepackage{dcolumn}
\usepackage{amsmath,amssymb}
\usepackage{bm}
\usepackage{color}
\usepackage{latexsym}
\usepackage{epstopdf}
\usepackage{color}
\usepackage[english]{babel}
\usepackage{latexsym}
\usepackage{psfrag,graphicx}
\usepackage{subfigure}
\usepackage{amsmath}
\usepackage{amssymb}
\usepackage{amsfonts}
\usepackage{bm}
\usepackage{natbib}
\usepackage{epstopdf}
\usepackage{appendix}

\definecolor{mygrey}{gray}{0.35}
\definecolor{myblue}{rgb}{0.2,0.2,0.8}
\definecolor{myzard}{cmyk}{0,0,0.05,0}
\definecolor{mywhite}{rgb}{1,1,1}
\definecolor{mywhite}{rgb}{1,1,1}
\definecolor{myred}{rgb}{1,0.,0.3}

\usepackage[colorlinks=true,citecolor=myblue,linkcolor=myred]{hyperref}

\def\ba{\begin{align}}
\def\enda{\end{align}}
\def\bi{\begin{itemize}}
\def\ei{\end{itemize}}

\def\be{\begin{equation}}
\def\ee{\end{equation}}
\def\bea{\begin{eqnarray}}
\def\eea{\end{eqnarray}}
\def\bse{\begin{subequations}}
\def\ese{\end{subequations}}

\begin{document}
\title{Quantum sensing of the phase space displacement parameters using a single trapped ion}
\author{Peter A. Ivanov}
\affiliation{Department of Physics, St. Kliment Ohridski University of Sofia, 5 James Bourchier blvd, 1164 Sofia, Bulgaria}
\author{Nikolay V. Vitanov}
\affiliation{Department of Physics, St. Kliment Ohridski University of Sofia, 5 James Bourchier blvd, 1164 Sofia, Bulgaria}

\begin{abstract}
We introduce a quantum sensing protocol for detecting the parameters characterizing the phase space displacement by using a single trapped ion as a quantum probe.
We show that thanks to the laser-induced coupling between the ion's internal states and the motion mode the estimation of the two conjugated parameters describing the displacement can be efficiently performed by a set of measurements of the atomic state populations.
Furthermore, we introduce a three-parameter protocol capable to detect the magnitude, the transverse direction and the phase of the displacement.
We characterize the uncertainty of the two- and three-parameter problems in terms of the Fisher information and show that state projective measurement saturates the fundamental quantum Cramer-Rao bound.
\end{abstract}


\maketitle

\section{Introduction}

Over the last few years the multiparameter estimation problems attract considerable interest in the light of its technological applications such as quantum-enhanced sensing and imaging.
Recently, the simultaneous enhanced estimation of multiple phases has been studied in photonic systems \cite{Humphreys2013,Baumgratz2016}.
Other examples include the joint estimation of phase and phase diffusion \cite{Vidrighin2014} as well as the multiple phase estimation in the presence of noise \cite{Yue2014}.

Among the other quantum systems, the trapped ions provide an excellent experimental platform with applications in high-precision quantum metrology but so far it has been mostly focussed on the estimation of a single parameter.
Examples include high-precision spectroscopy with multiparticle entangled states \cite{Leibfried2004}, sensing of the amplitude of motion \cite{Gilmore2017} as well as
highly sensitive detection of weak forces \cite{Knunz2010,Shaniv2017} and magnetic fields \cite{Kotler2014,Baumgart2016}.
However, many sensing protocols are inherently multi-parameter estimation problems since they involve the detection of the magnitude of the measured field and its phase.

In this work we propose a quantum sensing protocol for the detection of the parameters which characterize the phase space displacement operator by using a single trapped ion.
We discuss a quantum sensing protocol of two conjugated parameters, namely, the magnitude and the phase of the displacement by using a quantum probe consisting of three atomic states.
We show that thanks to the laser-induced coupling between the atomic states and the motion mode the two-parameter estimation can be efficiently carried out by state projective measurements in the original atomic-state basis.
Furthermore, we extend the sensing protocol by including the detection of the transverse direction of the displacement.
The three-parameter estimation can be performed by using a five-state system in four-pod configuration.
Measuring the atomic populations one can estimate the components of the force along two orthogonal directions and its phase.
We examine the sensitivity of the two- and three-parameter estimations in terms of the Fisher information matrices.
We show that the projective measurements in the original basis lead to an equality between the classical and quantum Fisher matrices and thus the uncertainty of the multiparameter estimation is bounded by the quantum Cramer-Rao bound.

The paper is structured as follows.
In Sec. \ref{BMS} we provide the general background on the multiparameter estimation.
In Sec. \ref{q_probe} we introduce the quantum probe represented by a single ion. 
In Sec. \ref{SP_two} we discuss the sensing protocol for the two parameters describing the phase space displacement.
In Sections \ref{TPE} and \ref{SP_3} we extend the sensing protocol to the estimation of three parameters including the magnitude, the transverse direction and the phase of unknown force.
Finally, in Sec. \ref{C} we summarize our findings.

\section{Background on the multiparameter estimation}\label{BMS}

Classical Fisher information (CFI) quantifies the amount of information on the parameters $\lambda=(\lambda_{1},\lambda_{2},\ldots,\lambda_{p})$
of a system that can be acquired for a given probe state and a specific set of measurement outcomes with probability $p_{n}(\lambda)$ with $n=1,2,\ldots,N$ \cite{Toth2014}.
For the multi-parameter estimation the CFI matrix elements read
\begin{equation}
I_{ij}=\sum_{n=1}^{N}\frac{1}{p_{n}}\left(\frac{\partial p_{n}}{\partial \lambda_{i}}\right)\left(\frac{\partial p_{n}}{\partial \lambda_{j}}\right).
\label{CFI}
\end{equation}
Defining the covariance matrix elements as $\Gamma_{ij}=\langle\lambda_{i}\lambda_{j}\rangle-\langle\lambda_{i}\rangle\langle\lambda_{j}\rangle$ 
 the following matrix inequality
\begin{equation}
\Gamma\geq (\nu \textbf{I})^{-1},\label{CCRB}
\end{equation}
is fulfilled, where $\nu$ is the number of the experimental repetitions.
The classical Fisher information is further bounded by the quantum Fisher information (QFI) matrix $\textbf{H}$ via the matrix inequality $\textbf{H}\geq\textbf{I}$, which gives the ultimate precision in the multi-parameter estimation quantified by the quantum Cramer-Rao bound
\begin{equation}
\Gamma\geq (\nu \textbf{H})^{-1}.\label{QCRB}
\end{equation}
The Symmetric Logarithmic Derivative (SLD) $\hat{L}_{\lambda_{j}}$ ($j=1,2,\ldots,p$) satisfies the operator equation
\begin{equation}
\frac{\partial \hat{\rho}}{\partial \lambda_{j}}=\frac{1}{2}\{\hat{L}_{\lambda_{j}}\hat{\rho}+\hat{\rho}\hat{L}_{\lambda_{j}}\},
\end{equation}
where $\hat{\rho}$ is the density operator, the corresponding QFI matrix elements are
\begin{equation}
H_{ij}=\frac{1}{2}{\rm Tr}\{\hat{\rho}(\hat{L}_{\lambda_{i}}\hat{L}_{\lambda_{j}}+\hat{L}_{\lambda_{j}}\hat{L}_{\lambda_{i}})\}.
\end{equation}

For a pure state we have $\hat{\rho}=\left|\psi\right\rangle\left\langle\psi\right|$ and the SLD operators simplify to $\hat{L}_{\lambda_{j}}=2\partial_{\lambda_{j}}\hat{\rho}$. Then the QFI matrix elements become
\begin{equation}
H_{ij}=4{\rm Re}\{\langle\partial_{\lambda_{i}}\psi|\partial_{\lambda_{j}}\psi\rangle-\langle\partial_{\lambda_{i}}\psi|\psi\rangle\langle\psi|
\partial_{\lambda_{j}}\psi\rangle\}.\label{QFI}
\end{equation}
The QFI is a measure of distinguishably of the quantum states with respect to the parameters of interest.
Indeed, it is straightforward to show that the infinitesimal distance between two adjusted states $|\psi_{\lambda}\rangle$ and $|\psi_{\lambda+d\lambda}\rangle$ can be defined by  $ds^{2}=2(1-|\langle\psi_{\lambda}|\psi_{\lambda+d\lambda\rangle}|)$.
Up to second order of the parameter variations the distance can be expressed as $ds^{2}=\sum_{i,j}g_{ij}d\lambda_{i}\lambda_{j}$ where the metric tensor is given by $g_{ij}=\frac{1}{4}H_{ij}$ \cite{Braunstein1994,Paris2009}.

Unlike the single-parameter estimation problem where the ultimate precision bound always is achieved in the basis of eigenvectors of the SLD operator, in multiparameter estimation the quantum Cramer-Rao bound is not always achievable since the SLD operators corresponding to different parameters may not commute.
Hence the ultimate precision for the two parameters is achieved by incompatible measurements.

A sufficient condition for the saturation of the quantum Cramer-Rao bound \eqref{QCRB} is the commutativity of the SLD operators on average, ${\rm Tr}\{\hat{\rho}[\hat{L}_{\lambda_{i}},\hat{L}_{\lambda_{j}}]\}=0$ \cite{Matsumoto2002}.
For a pure state this condition simplifies to
\begin{equation}
{\rm Im}{\langle\partial_{\lambda_{i}}\psi|\partial_{\lambda_{j}}\psi\rangle}=0,\quad (i,j=1,2,\ldots,p).\label{wcc}
\end{equation}

In the following we will discuss a quantum measurement protocol for detecting the parameters describing the phase-space bosonic-mode displacement.
We show that by using a quantum probe represented by a three-state system in the $\Lambda$ configuration driven by blue- and red-sideband laser
fields one can estimate the magnitude and the phase of the displacement operator simply by measuring the three atomic populations.
Furthermore, the additional information of the direction of the displacement can be acquired by considering a quantum probe represented by a five-state system in the four-pod configuration.
In both cases the estimation precision is bounded by Eq.~\eqref{QCRB}.

\section{Two-parameter estimation}\label{q_probe}

\subsection{Quantum probe}

In the following we consider a quantum probe represented by a single trapped ion with mass $M$ and trap frequencies $\omega_{x,y,z}$.
We assume that the atomic ion possesses three metastable internal electronic states $\left|m\right\rangle$ with $m=1,0,-1$ with atomic frequencies $\omega_{-1}$ and $\omega_{1}$ (we set the zero point energy to be $\omega_{0}=0$).
This is the case, for example, with the $^{171}$Yb$^+$ ion where the three-state system is formed by states $|F=0,m_{F}=0\rangle=|0\rangle$ and, respectively, $|F=1,m_{F}=\pm 1\rangle=\left|\pm 1\right\rangle$ \cite{Olmschenk2007}.
The goal of the present quantum sensing scheme is to estimate the two parameters $F$ and $\xi$ of the displacement operator
\begin{equation}
\hat{D}(F,\xi)=e^{F(\hat{a}^{\dag}e^{i\xi}-\hat{a}e^{-i\xi})},
\end{equation}
 where $\hat{a}^{\dag}$ and $\hat{a}$ are creation and annihilation operators of bosonic excitation corresponding to a harmonic oscillator of frequency $\omega_{z}$.
Single-parameter estimation of an unknown force with a single trapped ion was discussed in \cite{Maiwald2009,Ivanov2015,Ivanov2016}.
Here $\lambda_{1}=F$ and $\lambda_{2}=\xi$ are the parameters we wish to estimate which describe the magnitude and the phase of the force applied to the harmonic oscillator.
The effect of $\hat{D}(F,\xi)$ is to displace the amplitude of the ion's motional state acting on the vacuum state, such that the two parameters are encoded into the magnitude and the phase of the respective vibrational coherent state.
Consequently, the motional state tomography would allow one to extract information of both parameters.

Here we follow a different approach which utilizes the laser-induced coupling between the internal electronic states and the ion's vibrational mode which allows us to map the information of both parameters onto the atomic populations.
Since the CFI matrix is not invertible for $N=2$ and thus the estimation uncertainties are unbounded, the two-parameter estimation requires measurement with at least three outputs.
In the following, we consider the three atomic states as the elements of the positive operator-valued measure with $\sum_{\pm1,0}\hat{\Pi}_{m}=1$ where $\hat{\Pi}_{m}=|m\rangle\langle m|$ is the corresponding projective operator.

\begin{figure}
\includegraphics[width=0.45\textwidth]{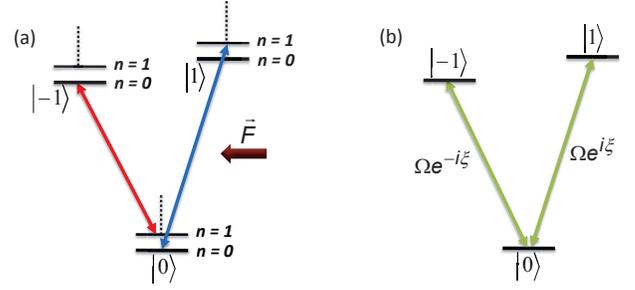}
\caption{(Color online) a) The probe system sensitive to the magnitude $F$ and the phase $\xi$ of the external applied force is represented by a
three-level system. We assume that the atomic transitions $\left|0\right\rangle\leftrightarrow\left|-1\right\rangle$ and $\left|0\right\rangle\leftrightarrow\left|1\right\rangle$
are driven respectively by red- and blue-sideband laser fields. b) For a time-varying force with oscillation frequency far from the resonance
with respect to the harmonic trap frequency the information of the two parameters is encoded into the magnitude and the phase of the Rabi frequencies
of the respective atomic transitions.}
\label{fig1}
\end{figure}

Consider that the ion interacts with two laser fields with frequencies $\omega_{{\rm L},1}$ and $\omega_{{\rm L},-1}$ applied along the trap axis $z$
which couple the two electronic states $\left|\pm1\right\rangle$ to state $\left|0\right\rangle$.
The total Hamiltonian describing the system is
\begin{eqnarray}
\hat{H}&=&\hbar\omega_{z}\hat{a}^{\dag}\hat{a}+\hbar\widetilde{\Omega}\{|1\rangle\langle0|e^{i\eta(\hat{a}^{\dag}+\hat{a})-i\delta_{1}t+i\varphi_{1}}
\notag\\
&&+|-1\rangle\langle0|e^{i\eta(\hat{a}^{\dag}+\hat{a})-i\delta_{-1}t+i\varphi_{-1}}+{\rm h.c.}\}+\hat{H}_{F},\label{H_total}
\end{eqnarray}
where $\widetilde{\Omega}$ is the Rabi frequency, $\eta=|\vec{k}|z_{0}\ll 1$ is the Lamb-Dicke parameter with $\vec{k}$ being the laser wave vector
pointing along the $z$ direction and $z_{0}=\sqrt{\frac{\hbar}{2M\omega_{z}}}$ is the spread of the axial oscillator ground-state wave function.
$\delta_{1}=\omega_{\rm{L},1}-\omega_{1}$ and $\delta_{-1}=\omega_{\rm{L},-1}-\omega_{-1}$ are the laser detunings and respectively $\varphi_{\pm1}$
are the laser phases.
The last term in Eq.~\eqref{H_total},
\begin{equation}
\hat{H}_{F}(t)=F z_{0}\cos(\omega_{d}t+\xi)(\hat{a}^{\dag}+\hat{a}),
\end{equation}
describes the effect of the external force with oscillation frequency $\omega_{d}$ which we assume to be $\omega_{d}=\omega_{z}-\omega$ where $\omega$ is the detuning ($\omega_{z}\gg\omega$).
The latter implies that only the vibrational mode along the trap axis is affected by the force such that the other two vibrational degrees of freedom can be neglected.
In the following we assume that the laser frequencies are tuned near the motional blue and red sidebands, $\omega_{\rm{L},1}=\omega_{1}+\omega_{z}-\omega-\Delta$ and  $\omega_{\rm{L},-1}=\omega_{-1}-\omega_{z}+\omega$.
The detuning $\omega$ introduces an effective phonon frequency along the trap axis, while the detuning $\Delta$ introduces an effective spin frequency on state $|1\rangle$, which can be used to compensate undesired AC Stark shifts \cite{Lee2005}.

We transform the Hamiltonian \eqref{H_total} into a rotating frame with respect to $\hat{U}_{R}(t)=e^{i\{\Delta|1\rangle\langle 1|-(\omega_{z}-\omega)\hat{a}^{\dag}\hat{a}\}t}$
and assume the Lamb-Dicke limit $\eta\sqrt{\langle \hat{a}^{\dag}\hat{a}\rangle+1}\ll1$, where $\langle \hat{a}^{\dag}\hat{a}\rangle$ is the average number of phonons.
Note that the unitary operator $\hat{U}_{R}$ commutes with the projective operators $\hat{\Pi}_{a}$ such that the measurement outcomes are not affected by the rotating-frame transformation.
By performing the vibrational rotating-wave approximation we arrive at the interaction Hamiltonian
\begin{subequations}\label{HI}
\begin{align}
\hat{H}_{I} &= \hat{H}_{0}+\hat{H}_{\rm sb},\\
\hat{H}_{0} &= \hbar\omega \hat{a}^{\dag}\hat{a}+\hbar\Delta|1\rangle\langle1|, \\
\hat{H}_{\rm sb} &= \hbar g\{\hat{a}^{\dag}\left|1\right\rangle\langle 0|+\hat{a}\left|-1\right\rangle\langle 0|+{\rm h.c.}\}
+\frac{Fz_{0}}{2}(\hat{a}^{\dag}e^{i\xi}+\hat{a}e^{-i\xi}),
\end{align}
\end{subequations}
where $\hat{H}_{I}=\hat{U}_{R}^{\dag}\hat{H}\hat{U}_{R}-i\hat{U}_{R}^{\dag}\partial_{t}\hat{U}_{R}$ and $g=\eta\tilde{\Omega}$ is the spin-phonon coupling.
In Eqs.~\eqref{HI} we have neglected the fast rotating terms which is valid as long as the conditions $|\omega_{z}-\omega|\gg \tilde{\Omega}$ and $2|\omega_{z}-\omega|\gg F z_{0}/2\hbar$ are fulfilled.
As a result of that the transition $\left|-1\right\rangle\leftrightarrow\left|0\right\rangle$ is driven by Jaynes-Cummings interaction, while the transition $\left|1\right\rangle\leftrightarrow\left|0\right\rangle$ by anti-Jaynes-Cummings interaction.
Note that by setting $g=0$ and assuming that the time-varying force is in resonance with respect to the harmonic trap frequency the time-evolution generated by $\hat{H}_{I}$ is simply given by the displacement operator $\hat{D}(F,\xi)$.
In that case the joint estimation of both parameters $F$ and $\xi$ describing the phase space displacement was discussed in \cite{Genoni2013}.
Here we use the laser-induced coupling between the three states and the motional mode as a mediator to map the information of both parameters directly on the respective three atomic populations.
As we will show below by proper choice of the initial atomic state the measurement strategy becomes optimal in a sense that it leads to the matrix equality $I_{ij}=H_{ij}$.

In the following we treat $\hat{H}_{\rm sb}$ in (\ref{HI}) as a perturbation term which is valid as long as the conditions $g\ll\omega$ and $F z_{0}/2\hbar\ll\omega$ are fulfilled.
Since the frequency $\omega$ defines the highest energy scale in the system the phonon excitations are highly suppressed which leads to simple three-state dynamics.
In order to trace out the vibrational degree of freedom we perform the unitary transformation $\hat{H}_{\rm eff}=e^{-\hat{S}}\hat{H}_{I}e^{\hat{S}}$, where the anti-Hermitian operator $\hat{S}$ is defined by the condition $\hat{H}_{\rm sb}+[\hat{H}_{0},\hat{S}]=0$, which yields
\begin{eqnarray}
\hat{S}=\frac{g}{\omega}[\hat{a}(\left|-1\right\rangle\langle0|+|0\rangle\left\langle1\right|)-{\rm h.c.}] + \frac{z_{0}F}{2\hbar\omega}(\hat{a}e^{-i\xi}-{\rm h.c.}).
\end{eqnarray}
In the lowest-order approximation the effective Hamiltonian becomes $\hat{H}_{\rm eff}=\hat{H}_{0}+\frac{1}{2}[\hat{H}_{\rm sb},\hat{S}]$ which gives
\begin{equation}
\hat{H}_{\rm eff}=\hbar\Delta|1\rangle\langle1|-\hbar\Omega\{|0\rangle\langle1|e^{i\xi}+|0\rangle\langle-1|e^{-i\xi}+{\rm h.c.}\}+\hat{H}_{\rm res}.
\label{Heff}
\end{equation}
Here $\Omega=z_{0}Fg/2\hbar\omega$ is the effective Rabi frequency which depends on the force magnitude $F$.
The residual interaction between the atomic states and the vibrational mode is quantified by
\begin{equation}
\hat{H}_{\rm res}=\hbar(g^{2}/\omega) [\hat{a}^{\dag}\hat{a}(|1\rangle\langle1|- |-1\rangle\langle-1|) + |1\rangle\langle1| ],
\end{equation}
where the last term can be compensated by setting the laser detining to $\Delta=-g^{2}/\omega$.
For an initial motional thermal state, the term $\hat{H}_{\rm res}$ would induces spin dephasing which limits the estimation precision.

In the following we consider the time evolution of the system under the Hamiltonian (\ref{Heff}).
We show that by a proper choice of the initial atomic state, the resulting state projective measurement of the atomic populations allows one to acquire information on both displacement parameters with estimation precision given by Eq.~\eqref{QCRB}.

\subsection{Sensing Protocol for $F$ and $\xi$}\label{SP_two}

Let us assume that the harmonic oscillator is prepared in the motional ground state, such that the term $\hat{H}_{\rm res}$ has no affect on the atomic-state evolution.
Then the generic atomic input state $\left|\psi_{0}\right\rangle$ evolves in time according to $\left|\psi(t)\right\rangle=\hat{U}(F,\xi)\left|\psi_{0}\right\rangle$, where $\hat{U}(F,\xi)$ is the unitary parameter-dependent transformation
\begin{equation}
\hat{U}(F,\xi) =\left[\begin{array}{ccc}
\frac{1}{2}(a+1) & \frac{1}{2}(a-1)e^{-2i\xi} & \frac{1}{\sqrt{2}}b e^{-i\xi} \\ \frac{1}{2}(a-1)e^{2i\xi}
& \frac{1}{2}(a+1) & \frac{1}{\sqrt{2}}b e^{i\xi} \\ -\frac{1}{\sqrt{2}}b^{*} e^{i\xi} &-\frac{1}{\sqrt{2}}b^{*} e^{-i\xi} & a^{*}
\end{array}\right].\label{U}
\end{equation}
Here $a$ and $b$ are the complex-valued Cayley-Klein parameters which in the case of exact resonance are $a=\cos\left(\frac{A}{2}\right)$ and $b=-i\sin\left(\frac{A}{2}\right)$ where $A=2\sqrt{2}\Omega t$ is the rms pulse area \cite{Ivanov2006}.

The measurement protocol starts by preparing the system in state $|\psi_{0}\rangle=|0\rangle$.
The system evolves for time $t^{\prime}$ according to the unitary propagator (\ref{U}).
Then a $\frac{\pi}{2}$ pulse is applied between states $\left|\pm1\right\rangle$, which creates the superposition state $\left|\pm1\right\rangle\rightarrow(\left|1\right\rangle\mp\left|-1\right\rangle)/\sqrt{2}$.
The resulting state vector $|\psi(t)\rangle=\hat{U}_{\frac{\pi}{2}}\hat{U}(F,\xi)|\psi_{0}\rangle$ is given by
\begin{eqnarray}
|\psi(t)\rangle&=&\sin(\sqrt{2}\Omega t)(\sin(\xi)\left|-1\right\rangle-i\cos(\xi)\left|1\right\rangle)\notag\\
&&+\cos(\sqrt{2}\Omega t)\left|0\right\rangle.
\label{psi}
\end{eqnarray}
It is straightforward to show that for the state vector (\ref{psi}) the SLD operators $\hat{L}_{F}$ and $\hat{L}_{\xi}$ do not commute.
On the other hand, the weak commutativity condition (\ref{wcc}) with $\lambda=\Omega,\xi$ is always satisfied and thus an optimal measurement that saturates the quantum Cramer-Rao bound exists.
The corresponding QFI reads
\begin{equation}
H_{ij} =\left[\begin{array}{cc}
8 t^{2} & 0  \\ 0
& 4\sin^{2}(\sqrt{2}\Omega t)
\end{array}\right],\label{F}
\end{equation}
which implies that the uncertainty in the joint estimation of the force magnitude and the phase is given by
\begin{equation}
\delta F\ge \frac{\hbar\omega}{\sqrt{2\nu}\,z_{0}g t},\quad \delta\xi\ge\frac{1}{2\sqrt{\nu}\,\sin(\sqrt{2}\Omega t)}.
\end{equation}
The optimal measurements that saturate the quantum Cramer-Rao bound were discussed in \cite{Pezze2017}.
Here the saturation of Eq. (\ref{QCRB}) can be achieved via projective measurement of the atomic populations with probability outcomes $p_{m}(F,\xi)=|\langle m|\psi\rangle|^{2}$ ($m=-1,0,1$).

\begin{figure}
\includegraphics[width=0.45\textwidth]{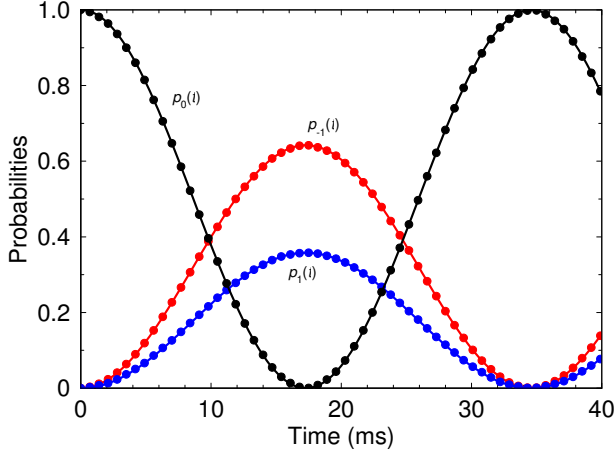}
\caption{(Color online)
The probabilities $p_{m}(t)$ versus time.
We compare the numerical solution of the time-dependent Schr\"odinger equation (circles) with the Hamiltonian \eqref{HI} to the analytical solution for the state vector \eqref{psi} (solid lines).
We set the laser detuning $\Delta=-g^{2}/\omega$ which compensates the AC-Stark shift.
The parameters are set to $g=4$ kHz, $\omega=150$ kHz, $F=-35$ yN, $\xi=1.7\pi$.
}
\label{fig2}
\end{figure}

In Fig.~\ref{fig2} we compare the analytical expressions for the probabilities with the exact results, where perfect agreement is observed.
From Eq.~\eqref{psi} it is straightforward to evaluate the corresponding CFI matrix which gives $H_{ij}=I_{ij}$.

\section{Three-parameter estimation}\label{TPE}

\subsection{Quantum probe}

Now we shall present an extension of our method to three-parameter estimation.
Consider that a time-varying force displaces the motional amplitude along the two orthogonal directions, so that
\begin{equation}
\hat{H}_{\vec{F}}(t)=\cos(\omega_{d}t+\xi)\{F_{x}r_{0}(\hat{a}_{x}^{\dag}+\hat{a}_{x})+F_{y}r_{0}(\hat{a}_{y}^{\dag}+\hat{a}_{y})\},
\end{equation}
where $\hat{a}^{\dag}_{x(y)}$ is the creation operator of vibrational quanta along the $x(y)$ direction.
Here $F_{x}$ and $F_{y}$ are the respective force components and $r_{0}=\sqrt{\frac{\hbar}{2M\omega_{\rm T}}}$ is the spread of the transverse oscillator ground state wave function with the trapped frequency $\omega_{x}=\omega_{y}=\omega_{\rm T}$.
Our goal is to estimate $\lambda=(F_{x},F_{y},\xi)$ via state projective measurements of the atomic populations.

\begin{figure}
\includegraphics[width=0.45\textwidth]{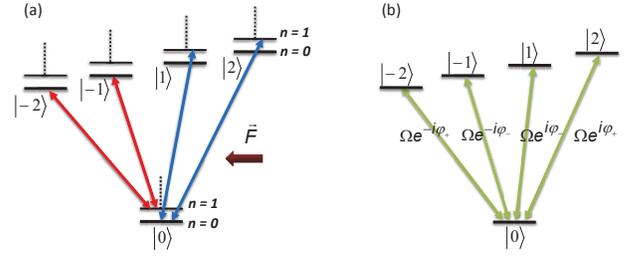}
\caption{(Color online)
(a) Probe system capable to detect the three parameters of the displacement operator, namely, the force components along the two orthogonal directions $F_{x}$ and $F_{y}$, and the phase $\xi$.
The quantum system consists of one ground state $\left|0\right\rangle$ and four excited states $\left|\pm1\right\rangle$ and $\left|\pm2\right\rangle$.
The red- and blue-detuned laser fields couple the internal states with the motional states along $x$ and $y$ directions.
(b) Adiabatic elimination of the phonon states leads to a closed set of states.
The information of the magnitude and the transverse direction of the force and its phase is encoded onto the magnitude of the Rabi frequency $\Omega$ and the phases $\varphi_{\pm}=\xi\pm\phi$.}
\label{fig3}
\end{figure}

In order to extract the information of the three parameters we assume that the atomic states are coupled to the vibrational states in both $x$ and $y$ directions via red- and blue-detuned laser fields with the laser configuration discussed above.
However, the measurement of the three atomic-state populations is not sufficient to determine all unknown parameters.
Indeed, the information of the transverse direction of the force is encoded in the phase in Eq.~\eqref{Heff} with the trivial redefinition $\xi\rightarrow \xi+\phi$ where we define $|\vec{F}_{\perp}|=\sqrt{F_{x}^{2}+F_{y}^{2}}$ and $\phi=\arctan\left(F_{y}/F_{x}\right)$.
Hence as long as the phase $\xi$ is known the estimations of the two force components can be carried out with the same approach as above.
However, for general three-parameter estimation problem we utilize five atomic states ($m=-2,\ldots,2$) as depicted in Fig.~\ref{fig3}, which are elements of the positive operator-valued measure with $\sum_{m=-2}^{2}\hat{\Pi}_{m}=1$.
In the following we show that by measuring the populations $p_{m}(t)$ one can determine all three parameters.

Consider a single trapped ion interacting with red- and blue-detuned laser fields which create Jaynes-Cummings and respectively anti-Jaynes-Cummings interactions between the atomic and vibrational states.
The interaction Hamiltonian in the Lamb-Dicke limit and after making the atomic and vibrational rotating-wave approximations is
\begin{equation}
\hat{H}_{I}=\hat{H}_{0}+\hat{H}_{x}+\hat{H}_{y}+\hat{H}_{\vec{F}},
\end{equation}
with
\begin{subequations}
\begin{align}
\hat{H}_{0}&=\hbar\omega(\hat{a}_{x}^{\dag}\hat{a}_{x}+\hat{a}_{y}^{\dag}\hat{a}_{y})+\hbar\Delta(|2\rangle\langle2|+|1\rangle\langle1|-|0\rangle\langle0|),\\
\hat{H}_{x}&=\hbar g\{a_{x}^{\dag}|0\rangle(\left\langle-1\right|+\left\langle-2\right|)+a_{x}^{\dag}(\left|1\right\rangle+\left|2\right\rangle)\langle0|+{\rm h.c.}\},\\
\hat{H}_{y}&=i\hbar g\{a_{y}^{\dag}|0\rangle(\left\langle-1\right|-\left\langle-2\right|)+a_{y}^{\dag}(\left|1\right\rangle-\left|2\right\rangle)\langle0|-{\rm h.c.}\},\\
\hat{H}_{\vec{F}}&=\frac{F_{x}r_{0}}{2}(\hat{a}^{\dag}_{x}e^{i\xi}+\hat{a}_{x}e^{-i\xi})+\frac{F_{y}r_{0}}{2}(\hat{a}^{\dag}_{y}e^{i\xi}+\hat{a}_{y}e^{-i\xi}).\label{Hxy}
\end{align}
\end{subequations}
Here the term $\hat{H}_{x}$ describes the Jaynes-Cummings transitions between the states $\left|-2\right\rangle\leftrightarrow\left|0\right\rangle$ and $\left|-1\right\rangle\leftrightarrow\left|0\right\rangle$, while the term $\hat{H}_{y}$ describes the respective anti-Jaynes-Cummings interaction between the states $\left|2\right\rangle\leftrightarrow\left|0\right\rangle$ and $\left|1\right\rangle\leftrightarrow\left|0\right\rangle$, see Fig. \ref{fig3}(a).
Both interactions couple the atomic states with the vibrational states in the two orthogonal $x$ and $y$ directions.
The detuning $\Delta$ is introduced to compensate the undesired AC Strak-shifts due to the interaction between the atomic and the vibrational states.
The last term $\hat{H}_{\vec{F}}$ describes the effect of the force, which creates a motional displaced states along the two directions with magnitudes proportional to $F_{x}$ and $F_{y}$ and phase $\xi$.
\begin{figure}
\includegraphics[width=0.45\textwidth]{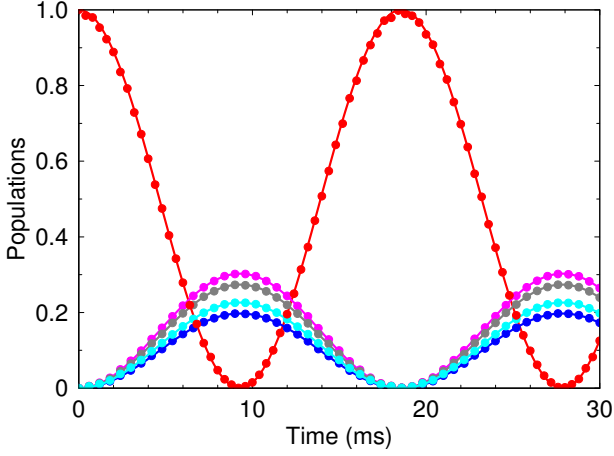}
\caption{(Color online)
The probabilities $p_{m}(t)$ versus time for the four-pod system.
We compare the numerical solution of the time-dependent Schr\"odinger equation (red circles) with the Hamiltonian \eqref{Hxy} after applying two $\pi/2$ pulses with the analytical solution for the state vector \eqref{psi_5} (solid lines).
We set the laser detuning $\Delta=-2g^{2}/\omega$, which compensates the AC-Stark shift.
The parameter as set to $g=4$ kHz, $\omega=150$ kHz, $F_{x}=-35$ yN, $F_{y}=-30$ yN, and $\xi=0.51\pi$.}
\label{fig4}
\end{figure}

In the weakly interacting regime, which is justified for $\omega\gg g$ and $\omega\gg F_{x,y}r_{0}/2\hbar$, one can adiabatically eliminate the phonon states by performing the unitary transformation of the Hamiltonian \eqref{Hxy} according to $\hat{H}_{\rm eff}=e^{-\hat{S}_{x}-\hat{S}_{y}}\hat{H}_{I}e^{\hat{S}_{x}+\hat{S}_{y}}$, with
\begin{subequations}
\begin{align}
\hat{S}_{x}&=\frac{g}{\omega}\{\hat{a}_{x}(\left|-1\right\rangle+\left|-2\right\rangle)\langle0|+\hat{a}_{x}|0\rangle(\langle1|+\langle2|)
-{\rm h.c.}\}\notag\\
&+\frac{F_{x}r_{0}}{2\hbar\omega}\hat{a}_{x}e^{-i\xi}-{\rm h.c.},\\
\hat{S}_{y}&=-i\frac{g}{\omega}\{\hat{a}_{y}(\left|-1\right\rangle-\left|-2\right\rangle)\langle0|+\hat{a}_{y}|0\rangle(\langle1|-\langle2|)
-{\rm h.c.}\}\notag\\
&+\frac{F_{y}r_{0}}{2\hbar\omega}\hat{a}_{y}e^{-i\xi}-{\rm h.c.}
\end{align}
\end{subequations}
To the lowest order of $g/\omega$ the effective Hamiltonian can be written as $\hat{H}_{\rm eff}=\hat{H}_{0}+\frac{1}{2}[\hat{H}_{x}+\hat{H}_{y},\hat{S}_{x}+\hat{S}_{y}]$, which yields
\begin{align}
\hat{H}_{\rm eff}&=-\hbar\Omega\{(e^{i\varphi_{+}}\left|-2\right\rangle+e^{i\varphi_{-}}\left|-1\right\rangle+e^{-i\varphi_{-}}
\left|1\right\rangle+e^{-i\varphi_{+}}\left|2\right\rangle)\langle0|\notag\\
&+{\rm h.c.}\}+\hat{H}_{\rm res},
\end{align}
where we set $\Delta=2g^{2}/\omega$ to compensate the undesired AC-Stark shifts of the atomic levels. Here $\Omega=|\vec{F}_{\perp}|g/2\hbar\omega$ is the Rabi frequency of the atomic transitions which is  proportional to the transverse force magnitude.
The two phases $\varphi_{\pm}=\xi\pm\phi$ encode the information of the transverse direction as well as the phase of the force.
The term $\hat{H}_{\rm res}$ describes the residual coupling between the atomic states and the vibrational modes and can be neglected as long as the both harmonic oscillators are prepared in the motional ground state.

\subsection{Sensing Protocol for $F_{x}$, $F_{y}$ and $\xi$}\label{SP_3}
The time-evolution of the five-state system is described by the parameter-dependent unitary matrix \cite{Ivanov2006}
\begin{widetext}
\begin{equation}
\hat{U}(F_{x},F_{y},\xi) =\left[\begin{array}{ccccc}
\frac{1}{4}(a+3) & \frac{1}{4}(a-1)e^{2i\phi} & \frac{1}{4}(a-1)e^{2i\xi} & \frac{1}{4}(a-1)e^{2i\varphi_{+}}& \frac{1}{2}be^{i\varphi_{+}} \\ \frac{1}{4}(a-1)e^{-2i\phi} & \frac{1}{4}(a+3) & \frac{1}{4}(a-1)e^{2i\varphi_{-}}& \frac{1}{4}(a-1)e^{2i\xi}&\frac{1}{2}be^{i\varphi_{-}} \\ \frac{1}{4}(a-1)e^{-2i\xi}& \frac{1}{4}(a-1)e^{-2i\varphi_{-}} & \frac{1}{4}(a+3)& \frac{1}{4}(a-1)e^{2i\phi_{-}}&\frac{1}{2}be^{-i\varphi_{-}}\\
\frac{1}{4}(a-1)e^{-2i\varphi_{+}}&\frac{1}{4}(a-1)e^{-2i\xi}&\frac{1}{4}(a-1)e^{-2i\phi}&\frac{1}{4}(a+3)&\frac{1}{2}be^{-i\varphi_{+}}\\
-\frac{1}{2}b^{*}e^{-i\varphi_{+}}&-\frac{1}{2}b^{*}e^{-i\varphi_{-}}&-\frac{1}{2}b^{*}e^{i\varphi_{-}}&-\frac{1}{2}b^{*}e^{i\varphi_{+}}&a^{*}
\end{array}\right], \label{UU}
\end{equation}
\end{widetext}
where $a=\cos(2\Omega t)$ and $b=-i\sin(2\Omega t)$.
The system is prepared initially in state $\psi(0)=|0\rangle$ and evolves in time according to Eq.~\eqref{UU}.
At time $t^{\prime}$ two $\pi/2$ pulses are applied between states $\left|\pm2\right\rangle$ and $\left|\pm1\right\rangle$, which create an equal superposition $\left|\pm1\right\rangle\rightarrow(\left|1\right\rangle\pm\left|-1\right\rangle)/\sqrt{2}$ and $\left|\pm2\right\rangle\rightarrow(\left|2\right\rangle\pm\left|-2\right\rangle)/\sqrt{2}$.
The state vector at time $t$ becomes
\begin{eqnarray}
|\psi(t)\rangle&=&\cos(2\Omega t)\left|0\right\rangle-\frac{i}{\sqrt{2}}\{\cos(\varphi_{+})\left|2\right\rangle+i\sin(\varphi_{+})\left|-2\right\rangle\notag\\
&&+\cos(\varphi_{-})\left|1\right\rangle+i\sin(\varphi_{-})\left|-1\right\rangle\}\label{psi_5}.
\end{eqnarray}

In Fig.~\ref{fig4} we compare the exact probabilities $p_{m}(t)$ ($m=-2,-1,\ldots,2$) with the probabilities obtained from the state vector \eqref{psi_5}.
Perfect agreement is observed.
It is straightforward to show that the necessary condition \eqref{wcc} for saturation of the fundamental bound \eqref{QCRB} with $\lambda=\Omega,\xi,\phi$ is always fulfilled for the state vector \eqref{psi_5}.
Again using Eq.~\eqref{psi_5} one can find the QFI matrix.
We obtain
\begin{equation}
H_{ij} =\left[\begin{array}{ccc}
4 t^{2} & 0 &0 \\ 0
& \sin^{2}(2\Omega t)&0\\
0&0&\sin^{2}(2\Omega t)
\end{array}\right],\label{F}
\end{equation}
such that the estimation uncertainty is bounded by
\begin{equation}
\delta F_{\perp}\ge\frac{\hbar\omega}{\sqrt{\nu}z_{0}g t},\quad \delta\xi=\delta\phi\ge\frac{1}{\sqrt{\nu}\sin(2\Omega t)}.\label{bound_xy}
\end{equation}
The estimation bounds \eqref{bound_xy} are achieved by state-projective measurements in the original atomic basis. Using Eq. (\ref{psi_5})
one can obtain the matrix equality $H_{ij}=I_{ij}$ which guarantees the saturation of the fundamental bound \cite{Pezze2017}.

Finally, we point out that as long as the laser fields driving the transitions $\left|\pm2\right\rangle\leftrightarrow\left|0\right\rangle$ are switch off then the sensing protocol is reduced to the detection of the magnitude and direction of the force with sensitivity $\delta F_{\perp}$ and $\delta\phi$ given by Eq. (\ref{bound_xy}).

\section{Conclusions}\label{C}

We have shown that a single ion can be used for estimating the parameters describing the phase-space displacement.
We have discussed the two-parameter estimation scheme using three internal ion's states driven by blue- and red- sideband laser fields.
We have shown that by measuring the respective atomic populations one can extract information about the two conjugated parameters, namely, the magnitude and the phase of the phase space displacement. Moreover, the sensing technique can be applied also for the estimation of the two component of the force.
We have extended the sensing protocol to three-parameter estimation including detection of the transverse direction, the magnitude and the phase of the measured force.
We have discussed the sensitivity of the multiparameter estimation problem in terms of the quantum Fisher information and we have shown that the projective measurement in the atomic basis saturates the fundamental quantum Cramer-Rao bound.


\acknowledgments
This work has been supported by the ERyQSenS project, Bulgarian Science Fund Grant No. DO02/3.

\end{document}